\documentclass[10pt]{article}
\usepackage[top=0.7in, bottom=0.7in, left=0.7in, right=0.7in]{geometry}
\usepackage[utf8]{inputenc}
\usepackage{caption,graphicx,subcaption, biblatex, amsmath, dirtytalk, float}
\usepackage{amssymb}
\usepackage{mathtools}
\usepackage{amsthm}
\usepackage{diagbox}
\usepackage[hidelinks]{hyperref}

\graphicspath{{./figures/}}
\DeclareUnicodeCharacter{0301}{\'{e}}

\usepackage{tablefootnote}
\usepackage{wrapfig} 

\usepackage[font=tiny]{caption}

\usepackage{placeins}

\title{\textsc{Unsupervised Learning of Phylogenetic Trees via Split-Weight Embedding}}
\author{
\textbf{Yibo Kong} \\
Department of Computer Science \\ 
University of Wisconsin-Madison\\
Madison, WI 53706
\and 
\textbf{George P. Tiley} \\
Kew Royal Botanic Gardens \\ 
Kew, Richmond, TW9 3AE, \\
London, UK
\and 
\textbf{Claudia Solís-Lemus}\thanks{Corresponding author: solislemus@wisc.edu}\\
Wisconsin Institute for Discovery\\
Department of Plant Pathology\\
University of Wisconsin-Madison\\
Madison, WI 53706
}

\begin{document}


\begin{center}
 \Large{Unsupervised Learning of Phylogenetic Trees via Split-Weight Embedding}  
 \begin{table}[ht]
     \centering
     \begin{tabular}{p{5cm}p{5cm}p{5cm}}
        Yibo Kong & George P. Tiley & Claudia Sol\'is-Lemus\tablefootnote{Corresponding author: solislemus@wisc.edu} \\
        University of Wisconsin-Madison & Kew Royal Botanic Gardens & University of Wisconsin-Madison
     \end{tabular}
 \end{table}
\end{center}

\begin{abstract}
\centering
\begin{minipage}{0.8\textwidth}

\noindent \textit{Motivation:} Unsupervised learning has become a staple in classical machine learning, successfully identifying clustering patterns in data across a broad range of domain applications. Surprisingly, despite its accuracy and elegant simplicity, unsupervised learning has not been sufficiently exploited in the realm of phylogenetic tree inference.
The main reason for application gap is the lack of a meaningful, yet simple, way of embedding phylogenetic trees into a vector space. Indeed, much work has been done in the embedding of genomic sequences, but embedding of graphical structures as phylogenetic trees for clustering tasks remains understudied.

\noindent \textit{Results:} Here, we propose the simple yet powerful split-weight embedding which embeds phylogenetic trees into a Euclidean space allowing us to fit standard clustering algorithms to samples of phylogenetic trees. Via extensive simulations and applications to real (\textit{Adansonia} baobabs) data, we show that our split-weight embedded clustering is able to recover meaningful evolutionary relationships. 

\noindent \textit{Availability and Implementation:}
We created the first open-source software for unsupervised learning of phylogenetic trees in the new Julia package \texttt{PhyloClustering.jl} available on GitHub \url{https://github.com/solislemuslab/PhyloClustering.jl}.

\noindent \textit{Contact:}
solislemus@wisc.edu.

\end{minipage}
\end{abstract}

\section{Introduction}
The Tree of Life is a massive graphical
structure which represents the evolutionary process from single cell organisms into the immense biodiversity of living species in
present time. 
Estimating the Tree
of Life would represent one of the greatest accomplishments in evolutionary biology and systematics, and it would allow scientists to understand the development and evolution of
important biological traits in nature, in particular, those related to
resilience to extinction when exposed to environmental threats such as
climate change. Therefore, the development of statistical and machine-learning theory
to reconstruct the Tree of Life, especially those scalable
to big data, are paramount in
evolutionary biology, systematics, and conservation efforts.

\begin{wrapfigure}{r}{7.6cm}
    \centering
    \includegraphics[width=7.6cm]{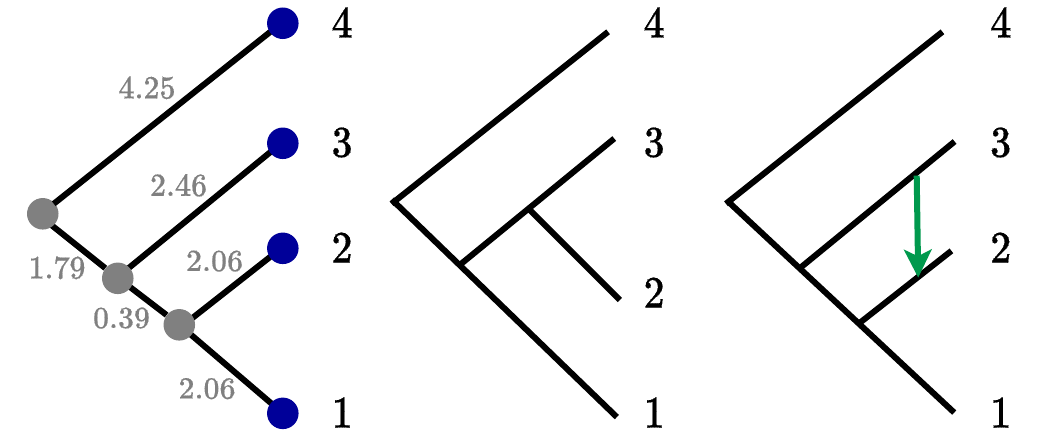}
    \caption{Left: Phylogenetic tree in 4 taxa. Internal (gray) nodes represent speciation events in which an ancestral species differentiates into two. External (blue) nodes, also called leaves, represent living species (here denoted 1,2,3,4). Edge weights (in gray) also called branch lengths can represent evolutionary time or expected substitutions per site. Center: Different phylogenetic tree on the same 4 taxa in which taxon 2 is grouped with taxon 3 rather than with taxon 1. Nodes are no longer drawn as is the most common representation of phylogenetic trees. Right: Phylogenetic tree with gene flow event depicted as a green arrow. This biological scenario represents the possibility that some genes have the evolutionary history of the phylogenetic tree on the left (with clade $(1,2)$) and some genes, the evolutionary history of the center tree (with clade $(2,3)$).}
    \label{fig:tree}
\end{wrapfigure}

The graphical structure
that represents the evolutionary process of a group of organisms is called a \textit{phylogenetic
tree}. A phylogenetic tree is a binary tree whose internal nodes represent ancestral species
that over time differentiate into two separate species giving rise to
its two children nodes (see Figure \ref{fig:tree} left). The evolutionary
process is then depicted by this bifurcating tree from the root (the
origin of life) to the external nodes of the tree (also called
leaves) which represent the living organisms today. Mathematically, a rooted phylogenetic tree $T$ on taxon set $X$ is a
connected directed acyclic graph with vertices
$V = \{r\} \cup V_L \cup V_T$ , edges $E$ and
a bijective leaf-labeling function $f : V_L \rightarrow X$ such that the root $r$ has indegree 0 and outdegree 2; 
any leaf $v \in V_L$ has indegree 1 and outdegree 0, and any internal node $v \in V_T$ has indegree 1 and outdegree 2. An unrooted tree results from the removal of the root node $r$ and the merging of the two edges leading to the outgroup (taxon 4 in Figure \ref{fig:tree} left).
Traditionally, phylogenetic trees are drawn without nodes (Figure \ref{fig:tree} center) given that only the bifurcating pattern is necessary to understand the evolutionary process. The specific bifurcating pattern (without edge weights) is called the tree topology. Edges in the tree have weight $w_e \in (0,\infty)$ that can represent different units, evolutionary time or expected substitutions per site being the most common. Trees with edge weights are called metric trees.

One of the main challenges when inferring the evolutionary history of species (denoted the species trees) is the fact that different genes in the data can have different evolutionary histories due to biological processes such as introgression, hybridization or horizontal gene transfer
\cite{introgression-drosophila, introgression-wild-tomatoes, introgression-humans}. An example is depicted in Figure \ref{fig:tree} (right) which has one gene flow event drawn as a green arrow. This gene flow event represents the biological scenario in which some genes in taxon 2 get transferred from the lineage of taxon 3, and thus, when reconstructing the evolutionary history of this group of four taxa, some genes will depict the phylogenetic tree that clusters taxa 1 and 2 in a clade (Figure \ref{fig:tree} left) and some genes will depict the phylogenetic tree that clusters taxa 2 and 3 in a clade (Figure \ref{fig:tree} center). Evolutionary biologists are interested in inferring the correct evolutionary history of the taxon group (species tree) accounting for the different evolutionary histories of individual genes.

In the absence of gene tree estimation error, the identification of the main trees that represent the evolution of certain gene groups would be trivial. However, estimation error combined with unaccounted biological processes such as incomplete lineage sorting or gene duplications complicate the identification of the main gene evolutionary histories in the data. Thus, novel and scalable tools to accurately classify gene trees that share common evolutionary patterns are needed.

Surprisingly, unsupervised learning methods have not fully been explored in phylogenetics. 
The only implementations of unsupervised learning in phylogenetics aim to cluster DNA sequences \cite{millan2022delucs, Ozminkowski2022-bw} or species \cite{derkarabetian2019demonstration, pyron2023unsupervised}, not trees.
The main challenge to implement unsupervised learning algorithms for phylogenetic trees is the discreteness of tree space. While there exists a proper distance function (Robinson-Foulds distance \cite{robinson1981comparison}) in the space of phylogenetic trees that can be used in clustering algorithms \cite{tahiri2022building}, the new trend in machine learning is to embed data points into an appropriately selected vector space \cite{vaswani2017attention, tenney2019bert, ji2021dnabert} to preserve meaningful similarity among vectors.
Traditionally, there are multiple natural embeddings designed for the space of phylogenetic trees. For example, the phylogenetic tree space introduced by Billera, Holmes, and Vogtmann (BHV space) \cite{billera2001geometry} is a continuous space for rooted $n$-taxon phylogenetic trees in which each tree is represented as a vector of edge weights in an orthant defined by its topology (bifurcating pattern). While the space is equipped in the geodesic distance, it is not a Euclidean space which can complicate the algorithms for the embedding and computation of distances. As a result, standard clustering algorithms, which rely on Euclidean geometry for efficiently measuring distances and defining cluster centroids, may not directly apply or need suboptimal adaptations to work within this complex, non-Euclidean framework. 
Other embeddings such as tropical geometric space \cite{monod2018tropical, lin2022tropical} or the hyperbolic embedding \cite{matsumoto2021novel, jiang2022learning} are mathematically sophisticated, yet unnecessarily complicated for standard clustering tasks.

Here, we define a simpler, yet equally powerful, embedding which we denote \textit{split-weight} which relies on the edge weights of taxon splits to embed an unrooted phylogenetic tree into a fully Euclidean vector space, and we implement three standard unsupervised algorithms (K-means \cite{kmeans}, Gaussian mixture model \cite{reynolds2009gaussian}, and hierarchical \cite{murtagh2012algorithms}) on the embedding space.
By testing the embedding on simulated and real data, we prove that it preserves phylogenetic tree similarity and allows us to cluster samples of gene trees into biologically meaningful groups. 
Furthermore, we present the first computational tool for the unsupervised learning of phylogenetic trees in the novel open-source Julia package \texttt{PhyloClustering.jl} available on GitHub \url{https://github.com/solislemuslab/PhyloClustering.jl} which is well-documented and user friendly for maximum outreach in the evolutionary biology. 

\section{Methods}
\subsection{Split-Weight Embedding of Phylogenetic Trees}
\label{sec:input}

\begin{wraptable}{r}{5cm}
\centering
    \begin{tabular}{ c|c|c }
    index & bipartition & $y(T)_i$ \\
    \hline
    0 & 1 $\mid$ 2 3 4 & 2.06 \\
    1 &	2 $\mid$ 1 3 4 & 2.06\\
    2 & 3 $\mid$ 1 2 4 & 2.46\\
    3 & 4 $\mid$ 1 2 3 & 6.04\\
    4 & 1 2 $\mid$ 3 4 & 0.39\\
    5 & 1 3 $\mid$ 2 4 & 0\\
    6 & 1 4 $\mid$ 2 3 & 0\\
    \hline
    \end{tabular} 
\caption{All possible bipartitions for 4 taxa ($\mathcal B_{\{1,2,3,4\}}$) with the numerical entry (third column) in the split-weight vector embedding for the unrooted version of the phylogenetic tree $T$ in Figure \ref{fig:tree} (left).}
\label{tab:bipartition}
\end{wraptable}

We define important notation in this section.

\noindent \textbf{Bipartitions.}
A bipartition (or \textit{split}) of the whole set of taxa ($X$) into two groups $X_1$ and $X_2$ is represented as $X_1 | X_2$ such that $X_1 \cap X_2 = \emptyset$ and $X_1 \cup X_2 = X$. Let $n=|X|$ denote the total number of taxa in the data. The number of bipartitions is given by $2^{n-1} - 1$. For example, for the case of $n=4$ taxa (Figure \ref{fig:tree}), there are 7 splits (Table \ref{tab:bipartition}).

\noindent \textbf{Split-Edge Equivalence.}
Let $\mathcal B_X$ be the set of all bipartitions for taxon set $X$ (with cardinality $2^{n-1} - 1$ as mentioned). Let $T$ be an unrooted phylogenetic tree on taxon set $X$. For every edge $e \in E$ in $T$, there exists a bipartition $b_e \in \mathcal B_X$ such that the removal of $e$ from the tree $T$ results in the two subsets of taxa defined by $b_e$. For example, the edge with weight $0.39$ in Figure \ref{fig:tree} (left) is an internal edge which, if removed from the tree, would split the taxa into two groups $X_1=\{1,2\}$ and $X_2=\{3,4\}$, and thus, this edge is uniquely mapped to the bipartition $b_e=12|34$. 
Thus, there is a mapping function $f_b: E \rightarrow \mathcal B_X$ such that $f_b(e)$ is the bipartition defined by $e$.
We note that not every bipartition is defined by an edge in $T$. For example, the split $13|24$ is not defined by any edge in the tree in Figure \ref{fig:tree} (left), and thus, the range of the mapping function $f_b(E) \subset \mathcal B_X$.

\noindent \textbf{Split Representation for Trees.}
Let $T$ be an unrooted $n$-taxon phylogenetic tree with taxon set $X$. For simplicity, instead of $f_b(E)$, we denote by $\mathcal B_T$ the bipartitions defined by $T$ as the splits $\{f_b(e)\}$ defined for every edge $e \in E$ in $T$. For example, the unrooted version of the phylogenetic tree in Figure \ref{fig:tree} (left) has five edges that define the bipartitions: $\mathcal B_T = \{1|234, 2|134, 12|34, 3|124, 4|123\}$ in post order traversal. Note that for the unrooted version of this tree, the two children edges of the root are merged into one edge with weight $6.04=1.79+4.25$.
In addition, we define a mapping function $f_w: \mathcal B_T \rightarrow \mathbb (0,\infty)$ that assigns a numerical value to each bipartition which corresponds to the edge weight of the uniquely mapped edge in the tree. For example, $f_w(12|34) = 0.39$.

\noindent \textbf{Split-Weight Embedding.}
Let $T$ be an $n$-taxon unrooted phylogenetic tree on taxon set $X$. Let $\mathcal B_T$ be the bipartitions defined by $T$, and let $\mathcal B_X$ be the set of all bipartitions of $X$. We define the split-weight embedding of $T$ as a ($2^{n-1} - 1$)--dimensional numerical vector $y(T) \in \mathbb [0,\infty)^{2^{n-1} - 1}$ such that the $i$th element of $y(T)$ corresponds to the $i$th bipartition ($b_i$) in $\mathcal B_X$ with entry value given by:
\begin{align*}
y(T)_i = \left\{ \begin{array}{ll}
f_w(f_b(e)) & if \ f_b^{-1}(b_i) = e \in E  \\
0 & otherwise
\end{array} \right.,
\end{align*}
That is, if there is an edge $e \in E$ in $T$ such that $f_b(e) = b_i$, then the $i$th entry of $y(T)$ is given by the corresponding edge weight $f_w(f_b(e))$. For example, for 4 taxa, there are 7 bipartitions; five of them correspond to edges in the tree in Figure \ref{fig:tree} left (Table \ref{tab:bipartition}) and for those, their value in the split-weight embedding vector correspond to the edge weights in $T$. In contrast, the two bipartitions that do not belong to $\mathcal B_T$ ($13|24$ and $14|23$) have a value in the split-weight embedding vector of $0$, as in \cite{kuhner}.
Thus, the split-weight embedding for the unrooted version of the phylogenetic tree $T$ in Figure \ref{fig:tree} (left) is the numerical vector $y(T)=(2.06, 2.06, 2.46, 6.04, 0.39, 0, 0)$.



\subsection{Clustering Algorithms}
\begin{figure}[ht]
    \begin{center}
\includegraphics[width=0.9\textwidth]{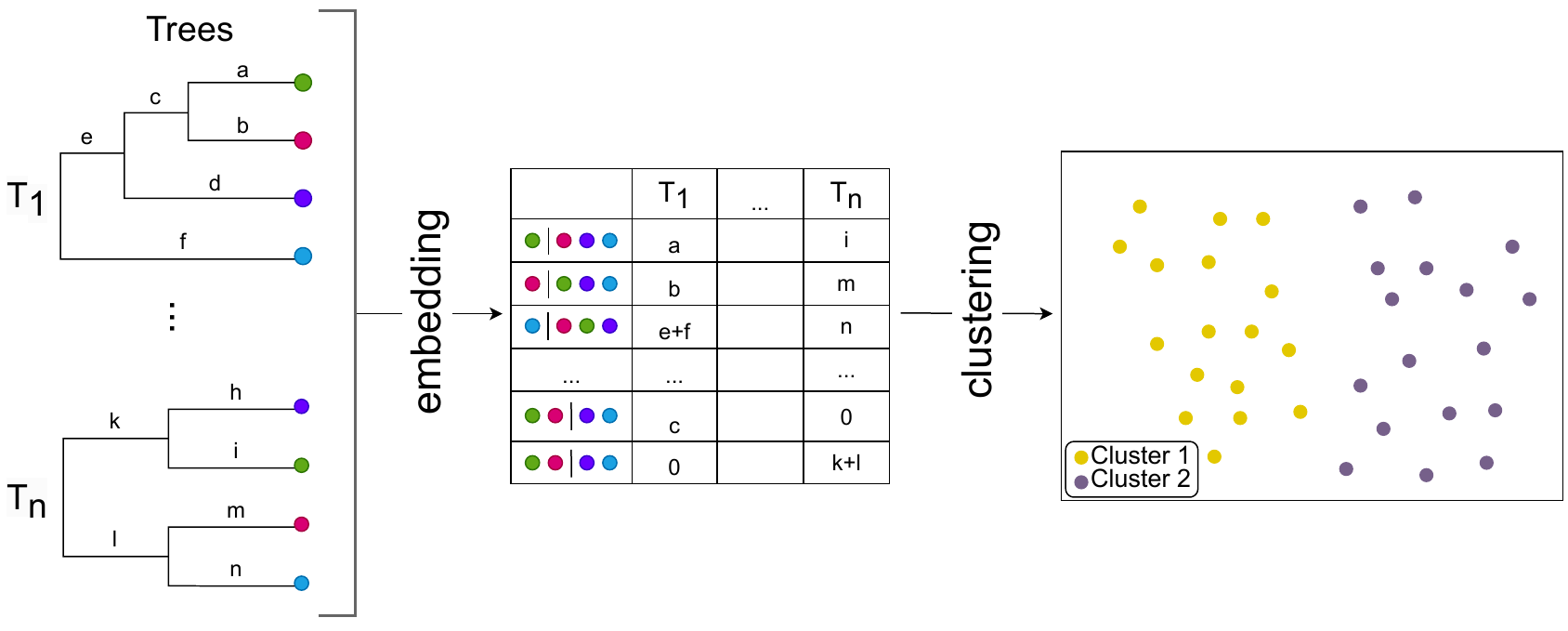}
    \caption{Graphical representation of our unsupervised learning algorithm for phylogenetic trees. In this case, we are embedding 4-taxon trees into a Euclidean space using the split-weight embedding to fit unsupervised learning models. Each point in the scatter plot is a tree. The dimension reduction strategy we use in this case for visualization is principal component analysis (PCA).}
    \label{fig:abstrct}
\end{center}
\end{figure}

For a sample of $N$ $n$-taxon unrooted phylogenetic trees $\{T_1,...,T_N\}$, we first embed them into the split-weight numerical vectors: $\{y(T_1),...,y(T_N)\}$. 
Note that all vectors have the same dimension as all the trees have the same taxon set ($X$).
The input matrix ($M \in [0,\infty)^{(2^{n-1} - 1) \times N}$) for unsupervised learning algorithms becomes the concatenated embedded vectors.
Then, we implement three different types of unsupervised learning methods which are modified to use the input matrix and directly output the predicted labels for each tree in the sample. Figure \ref{fig:abstrct} shows a graphical representation of our unsupervised learning strategy.

\noindent \textbf{K-means.} 
The K-means algorithm \cite{kmeans} is a partitioning method that divides data into \textit{K} distinct, non-overlapping clusters based on their characteristics, with each cluster represented by the mean (centroid) of its members. It starts by selecting \textit{K} initial centroids and iterates through two main steps: assigning data points to the nearest centroids, and then updating the position of the centroids to the mean of their assigned points. The algorithm minimizes the squared Euclidean distances within each cluster and ends when the positions of the centroids no longer move. 
Instead of traditional K-means, we use Yinyang K-means \cite{yinyang} which has less runtime and memory usage on large datasets. Traditional K-means calculates the distance from all data points to centroids for each iteration. Yinyang K-means uses triangular inequalities to construct and maintain upper and lower bounds on the distances of data points from the centroids, with global and local filtering to minimize unneeded calculations.

The parameters of this algorithm include the number of clusters ($K$), the initialization method, and the random seed for initialization. To select the initial centroids efficiently, we use the K-means++ \cite{plusplus} as default initializing algorithm.  
K-means++ initializes centroids by first randomly selecting one data point as a centroid, and then iteratively choosing subsequent centroids from the remaining points with probabilities proportional to the square of their distances from the nearest existing centroid.  The resulting centroids are typically spread out across the entire dataset. 

We also employ the repeating strategy \cite{repkmeans} for large datasets to mimic the behavior of a user who runs the algorithm multiple times and selects the most reasonable data in real-world situations. Meanwhile, this strategy avoids the instability of standard K-means clustering algorithm due to the initial centroid selection. This instability can easily result into convergence to a local minimum, even with the help of a better initial point selection algorithm like K-means++. 
Therefore, we use a simple strategy of repeating the K-means clustering at least 10 times and retaining the result with the highest accuracy when calculating the accuracy for each large group in our simulations. 


  


\noindent \textbf{Gaussian mixture model (GMM).} 
The GMM algorithm \cite{Buitinck2013-dn} is a model-based probabilistic method that assumes data points are generated from a mixture of several Gaussian distributions with unknown parameters. It uses the Expectation Maximization (EM) algorithm to update the parameters iteratively in order to optimize the log-likelihood of the data until convergence. The choice of covariance type in the GMM shapes how clusters are formed by allowing for different levels of feature interaction. This choice impacts the model's accuracy and its ability to generalize to new data. It is similar yet more flexible than K-means by permitting mixed membership of data points among clusters.
The parameters of this model include the number of clusters, the initialization method, and the covariance type. We use K-means as the method for selecting initial centrals and choose diagonal covariance as covariance type. We also employ repeating K-means \cite{repkmeans} to find the best starting centers.



\noindent \textbf{Hierarchical clustering.} 
The algorithm \cite{murtagh2012algorithms} is a hierarchical method that creates a clustering tree called a dendrogram. Each leaf on the tree is a data point, and branches represent the joining of clusters. We can `cut' the tree at different heights to get a different number of clusters. This method does not require the number of clusters to be specified in advance and can be either agglomerative (bottom-up) or divisive (top-down).
The parameters of this model are the linkage method, and the number of clusters. The linkage method in hierarchical clustering defines the metric used to compute the distance between observed datasets, which directly affects the shape and size of the clusters. As default, we use Ward's method \cite{ward} as linkage method for hierarchical clustering.


\subsection{Simulation Study}

We describe specific simulation scenarios in more details in the following subsections, but as an overall description: we perform simulation tests of K-means, GMM, and hierarchical clustering with \textit{K} = 2 for samples of $2N$ trees, each tree with $n$ taxa where we vary $N$ and $n$ as described below. For a given combination of $n$ and $N$, we select two species trees to serve as the theoretical centroids of each cluster. We generate 100 datasets to account for performance variability.
Tree variability within a cluster is simulated with the \texttt{PhyloCoalSimulations} 
Julia package \cite{simulate} which takes a species tree as input and simulates gene trees under the coalescent model \cite{rannala2003bayes} which generates tree variability due to random sorting of lineages within populations. In this sense, we expect simulated trees to cluster around the chosen species tree. The level of expected variability in the sample of gene trees is governed by the edge weights in the species tree (longer branches resulting in less tree variability) \cite{degnan2009gene}.
All trees are embedded into the split-weight space, and the vectors are normalized (i.e., set the mean to 0 and the standard deviation to 1)  prior to clustering. Visualization of clusters is performed via PCA.
To quantify the performance of unsupervised learning models, we use the Hungarian algorithm \cite{munkres} to find the grouping method with the largest overlap between predicted labels and true labels. In this grouping method, the fact that the predicted label is different from the true label means that the tree is not classified into the correct group. Therefore, the accuracy is defined as $\frac{2N -  t}{2N}$
where $t$ is the number of trees with different predicted and true labels. Higher accuracy means that unsupervised learning models are able to capture simulated trees originating from the same species tree more successfully.

\noindent \textbf{Clustering of trees with different topologies.}
For $n=4$ taxa, there are 15 possible bifurcating patterns (tree topologies).
For each of the 15 4-taxon species trees, 
we simulated samples of $N=50, 100, 500, 1000, 5000$ trees with \texttt{PhyloCoalSimulations} \cite{simulate} for two choices of edge weights in the species trees: 1) all edge weights set to $1.0$ coalescent unit, or 2) each edge weight randomly selected from an uniform distribution $(0.5,2)$ coalescent units. The rationale is that $1.0$ coalescent unit generates \textit{medium} levels of gene tree discordance \cite{degnan2009gene}, and thus, we expect the clustering algorithms to perform accurately, whereas shorter branches (e.g. $0.5$ coalescent units) produce more tree heterogeneity further complicating clustering.

For the case of more than 4 taxa, we cannot list all of the tree topologies (10,395 total unrooted trees for 8 taxa and 213,458,046,676,875 total unrooted trees for 16 taxa), so we randomly generate 15 8-taxon trees and 15 16-taxon trees using the simulating algorithm in the R package \texttt{SiPhyNetwork} \cite{siphynetwork} under a birth-death model. We focus on the case of edge weights randomly chosen uniformly in the interval $(0.5,2)$.

\noindent \textbf{Clustering of trees in NNI-neighborhoods.}
The Nearest Neighbor Interchange (NNI) move is a type of phylogenetic tree arrangement that selects an internal branch of a given tree and then swaps adjacent subtrees across that branch. It generates alternative tree topologies that are ``nearest neighbours'' to the original tree, differing only in the local arrangement of the subtrees connected by the chosen branch. For a given tree, we can define a NNI-neighborhood as all the trees that are one NNI move away from the selected tree (or any number of NNI moves away). In this section, we test whether the clustering algorithms are accurate enough to distinguish trees within the same NNI-neighborhood.
Using the simulating algorithm in the R package \texttt{SiPhyNetwork} \cite{siphynetwork} under a birth-death model, we randomly generate one 8-taxon species tree and one 16-taxon species tree, and then we perform 1, 2, 3, 4 and 5 NNI moves on each tree to produce 10 new trees (5 with 8 taxa and 5 with 16 taxa). The NNI function for phylogenetic trees is implemented in \texttt{PhyloNetworks} \cite{phylonetworks}. The edge weights are randomly selected from an uniform distribution $(0.5,2)$ coalescent units.
For each of the species trees, 
we simulate samples of $N=50, 100, 500, 1000, 5000$ trees with \texttt{PhyloCoalSimulations} \cite{simulate}.

\noindent \textbf{Clustering of trees with the same topology, but different edge weights.}
To test the performance of the clustering algorithms to classify trees that have the same topologies, but different edge weights, we simulate trees under the same species tree topology with six different sets of edge weights: 1) all edge weights equal to $1$ (denoted \textit{org} in the figures); 2) all edge weights lengths equal to $1.5$ (denoted \textit{inc} in the figures); 3) randomly selected edge weights uniformly in $(1, 2)$ (denoted \textit{inc\_r} in the figures); 4) all branch lengths equal to $0.5$ (denoted \textit{dec} in the figures); 5) randomly selected edge weights uniformly in $(0, 1)$ (denoted \textit{dec\_r} in the figures), and 6) randomly selected edge weights uniformly in $(0, 2)$ (denoted \textit{all\_r} in the figures). We only focus on the 4-taxon tree topologies for these tests.
For each of the species trees, 
we simulated samples of $N=50, 100, 500, 1000, 5000$ trees with \texttt{PhyloCoalSimulations} \cite{simulate}.

\noindent \textbf{Clustering performance when clusters of trees are imbalanced.}
We also test the clustering algorithms on imbalanced datasets with species trees with $n=4$, and $n=8$ taxa. For a given taxon size ($n$), we randomly generate two species tree, and for each species tree, we simulate $N$ trees with \texttt{PhyloCoalSimulations} \cite{simulate}. In one scenario, one of the clusters will have $N=100$ trees, and the other will have $N=1000$ trees (denoted $2N=1100$ in the Results). In the other scenario, one of the clusters will have $N=1000$ trees, and the other will have $N=6000$ trees (denoted $2N=6000$ in the Results).
Since our naive definition of accuracy cannot work on imbalanced data, we employ the Hubert-Arabie adjusted Rand index (ARI)\cite{Steinley2004PropertiesOT} to quantify the performance of clustering algorithms. The ARI is a refined version of the Rand index \cite{Hubert_Arabie_1985} that adjusts for the chance grouping of elements, thus providing a more accurate measure of the similarity between two clusterings by considering both the agreement and disagreement between clusters. An ARI of 1 indicates that the clustering results are perfectly identical, while an ARI of -1 signifies that the clustering results are entirely dissimilar. A score close to 0 suggests that the clustering outcomes are largely random. 

\noindent \textbf{Comparison of clustering from split-weight embeddings to Robinson-Foulds tree distances.}
As mentioned in the Introduction, clustering could be performed using pairwise tree distances from a inherent distance function on trees. The most widely used tree distance is Robinson-Foulds distance \cite{robinson1981comparison}.
We compare the performance (in terms of accuracy, time and memory) of the clustering task from split-weight embeddings vs using Robinson-Foulds distance on the samples of trees directly.
We choose trees with $n = 4, 8$ taxa, and clusters of size $N = 1000$. For this comparison, we focus on hierarchical clustering.

\subsection{Reticulate evolution in baobabs}

In-frame codon alignments of baobab target-enrichment data \cite{baobabs} are used to estimate gene trees under maximum likelihood (ML) \cite{felsenstein1981evolutionary} with \texttt{IQTREE} v.2.1.3 and default settings \cite{minh2020iq}. The ML analyses treats alignments as nucleotide data and the best model is determined by \texttt{ModelFinderPlus} \cite{kalyaanamoorthy10fast}, which uses the Bayesian Information Criterion for model selection. 
The data then consist of 372 estimated trees in 8 species of \textit{Adansonia}:
\textit{A. digitata} (continental Africa),
\textit{A. gregorii} (Australia),
\textit{A. grandidieri} (Madagascar),
\textit{A. suarezensis} (Madagascar),
\textit{A. madagascariensis} (Madagascar),
\textit{A. perrieri} (Madagascar),
\textit{A. za} (Madagascar), and
\textit{A. rubrostipa} (Madagascar). The data also contain an outgroup \textit{Scleronema micranthaum}, so in total, there are 9 taxa.
We remove 4 trees that only had 8 taxa, and 5 outlier trees with pathologically long edges so that the final dataset contain 363 trees which we embed in the split-weight space. We standarize the resulting matrix as in the simulation study, and cluster the vectors using the K-means algorithm ($K=2$). 
After clustering, we use Densitree \cite{densitree} to identify the consensus tree of each cluster via the root canal method.

\section{Results}

\subsection{Simulation Study}

\begin{figure*}
    \centering
    \includegraphics[scale=0.14]{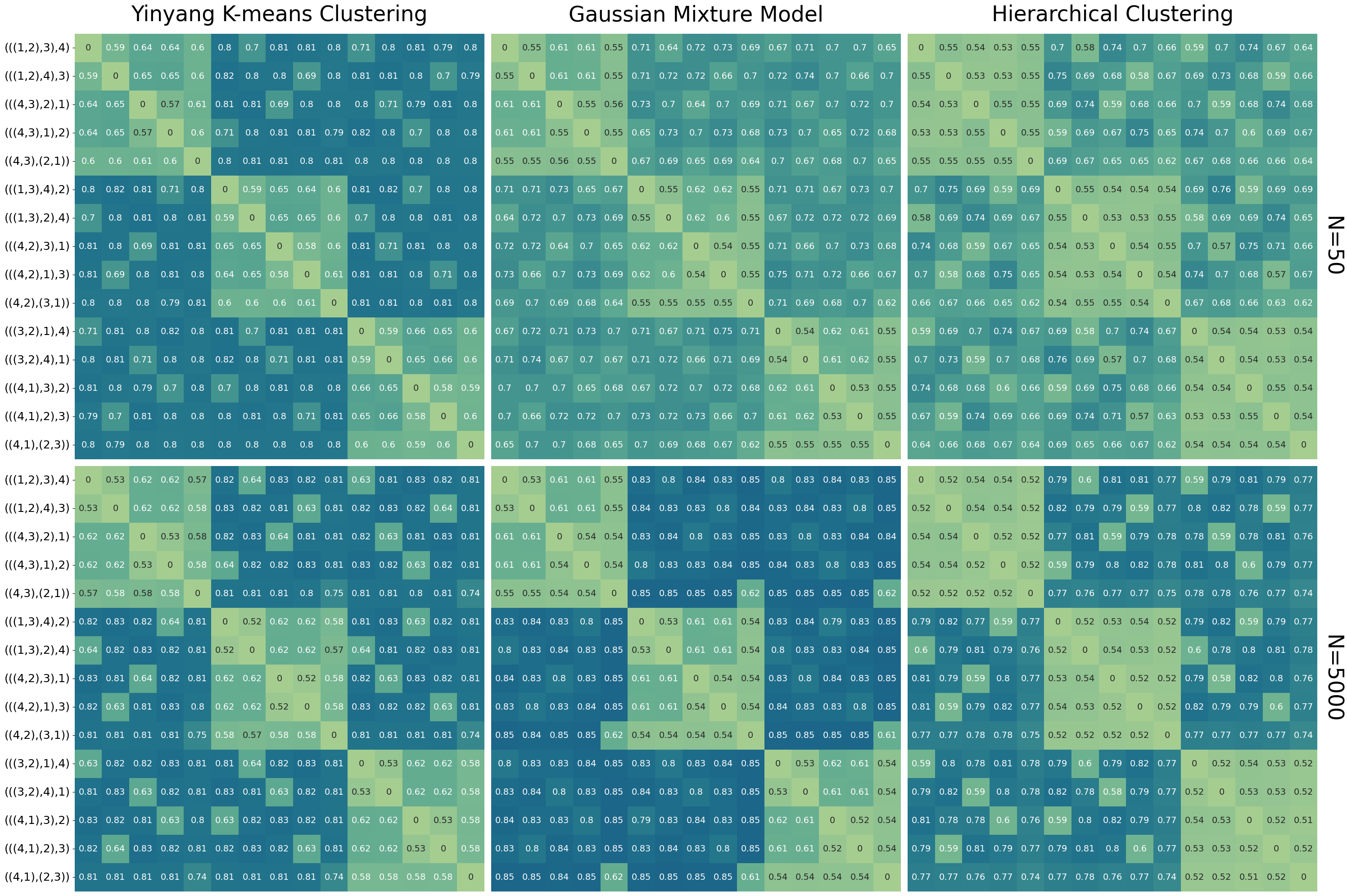}
    \includegraphics[scale=0.14]{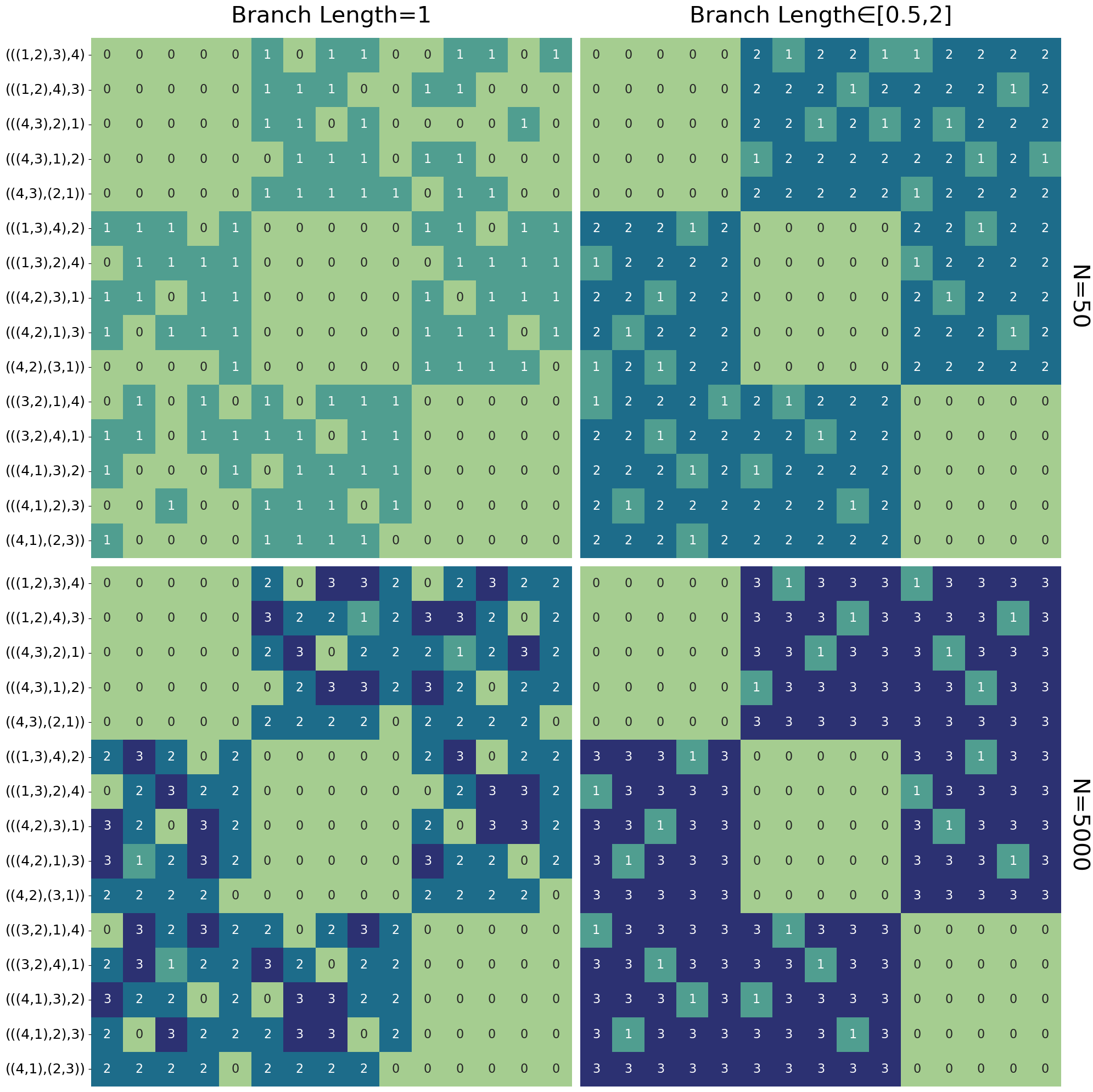}
    \caption{Left: Heatmaps of classificacion accuracy on the 15 trees with 4 taxa. Each panel corresponds to the classification accuracy of pairwise comparisons for two clusters originated on two trees (row and column) for one clustering method (columns: K-means, GMM and Hierarchical) and one sample size (rows: $N=50$ and $N=5000$). Within each panel, trees are sorted based on their unrooted topology (first 5 rows correspond to split $12|34$; next 5 to the split $13|24$, and last 5 to the split $14|23$). Off-diagonal blocks (trees with different unrooted representation) can be accurately separated as different clusters by GMM with large sample size ($N=5000$), and by K-means for any sample size. Right: Heatmaps on the number of algorithms that reach $80$\% classification accuracy ($0,1,2,$ or $3$ algorithms) by edge weight strategy (columns: edge weights equal to $1$ or edge weights randomly chosen in $(0.5,2)$) and sample size (rows: $N=50, 5000$).}
    \label{fig:4taxa}
\end{figure*}

Figure \ref{fig:4taxa} (left) shows a heatmap of the prediction accuracy of the different clustering algorithms to assess the performance of clustering of trees with different topologies for $n=4$ taxa. Each cell in the heatmap represents a comparison between the row and column tree (only rows labeled, but the order of the columns is the same). Trees are ordered depending of their unrooted topology. For example, the first 5 rows (and columns) correspond to the 5 different rooted versions of the unrooted tree corresponding to the split $12|34$; the next 5 rows (and columns) to the split $13|24$, and the last 5 rows (and columns) to the split $14|23$. The darker the color, the more accurate the classification of the two trees. We can see a diagonal block pattern in the heatmaps which illustrates the difficulty of separating two clusters defined by two rooted trees with the same unrooted representation. The heatmaps are arranged by clustering algorithm (three columns: K-means, GMM and hierarchical) and number of trees in the sample ($N=50$ and $N=5000$). Similar plot for $N=100,500,1000$ can be found in the Appendix (Figure \ref{fig:4taxab}). We can see that the accuracy of K-means is robust to sample size, while the accuracy of GMM is higher for larger number of trees ($N=5000$). Hierarchical clustering shows the worst performance of the three methods.

Figure \ref{fig:4taxa} (right) shows a different type of heatmap in which we summarize the performance of all three methods. Each cell in the heatmap can have four values: $0$ if none of the three methods have a classification accuracy above $80$\%; $1$ if one of the three methods has a classification accuracy above $80$\%; $2$ if two of the three methods have a classification accuracy above $80$\%, and $3$ if all three methods have a classification accuracy above $80$\%. The two columns correspond to the strategy to set edge weights (all edge weights equal to $1.0$ in the left column and edge weights randomly selected in $(0.5,2)$ in the right column). Unlike standard phylogenetic methods that tend to perform better when edges are $1.0$ coalescent unit long, the clustering methods here tested perform better with variable edge weights. Furthermore, we can see that with larger sample sizes ($N=5000$), most methods are able to distinguish samples originated from trees that do not have the same unrooted topology (off-diagonal blocks).

Figures \ref{fig:8taxa} and \ref{fig:16taxa} in the Appendix show the classification of all methods to cluster samples of trees for $n=8$ taxa and $n=16$ taxa respectively. 
%
The results for NNI-neighborhoods (Figures \ref{fig:nni} and \ref{fig:nnib}), same topology (Figures \ref{fig:bl1}, \ref{fig:bl2} and \ref{fig:bl3}) and imbalanced number of trees per cluster (Table \ref{tab:imbalanced}) are also in the Appendix. In all cases, the split-weight embedding is able to accurately cluster samples of trees.

Table \ref{tab:cost} shows the results of the comparison with the Robinson-Foulds hierarchical clustering. While accuracy is comparable, the split-weight embedding requires less time and less memory.

\begin{table}[h]
    \centering
        \begin{tabular}{ |c|c|c|c|c| } \hline
         & $n$ & Running time (seconds) & Memory cost (GiB) & Accuracy \\
        \hline
        Split-weight & 4 & 1.906 & 0.850 & 0.942 \\ \hline
        Robinson-Foulds & 4 & 11.193 & 5.295 & 0.927 \\ \hline
        Split-weight & 8 & 5.929 & 4.572 & 0.996 \\ \hline
        Robinson-Foulds & 8 & 37.146 & 25.557 & 0.999 \\
        \hline
        \end{tabular} 
\caption{Running time in seconds, memory cost in gibibytes (GiB), and accuracy for hierarchical clustering under two types of distances: distance from the split-weight embeddings and Robinson-Foulds distance on trees directly.}
\label{tab:cost}
\end{table}

\subsection{Reticulate evolution in baobabs}
\label{sec:canis}

\begin{figure*}[ht]
\vskip 0.2in
\begin{center}
\centerline{
\includegraphics[scale=0.16]{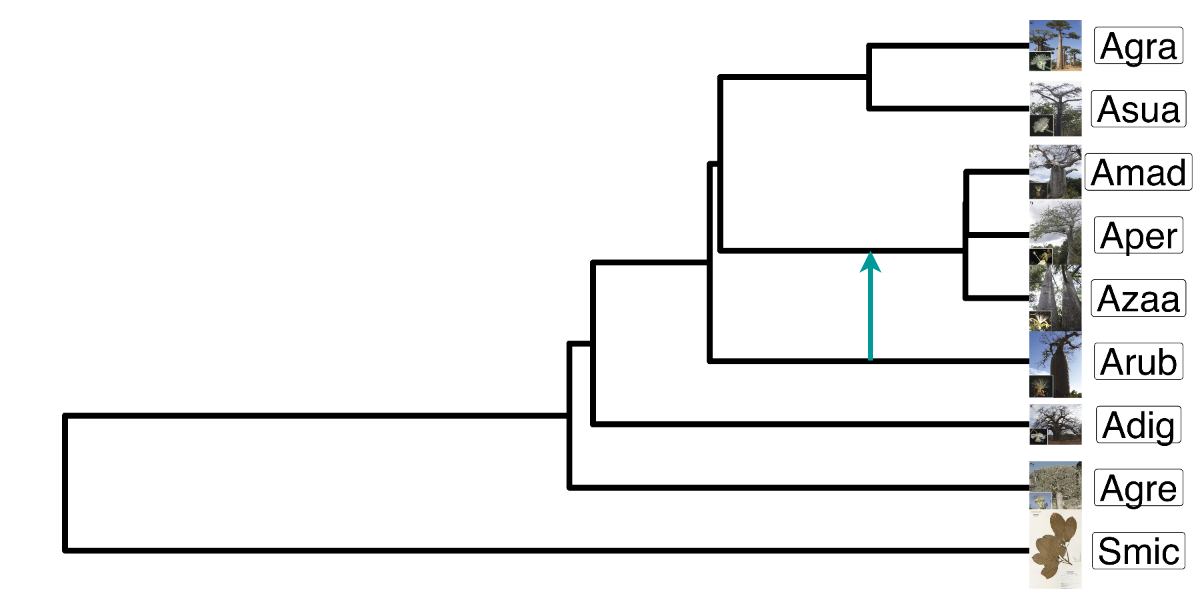}
\includegraphics[scale=0.03]{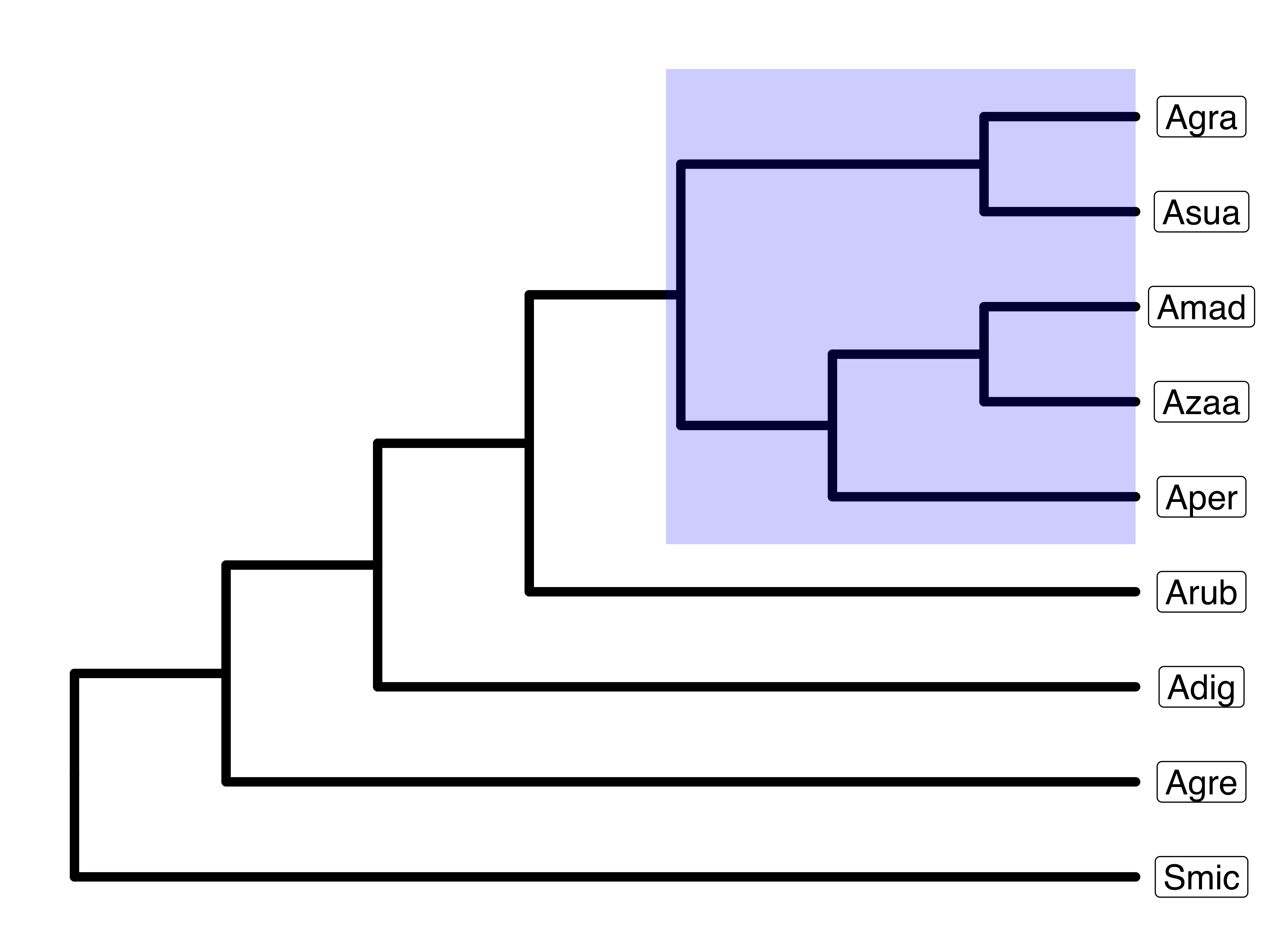}
\includegraphics[scale=0.03]{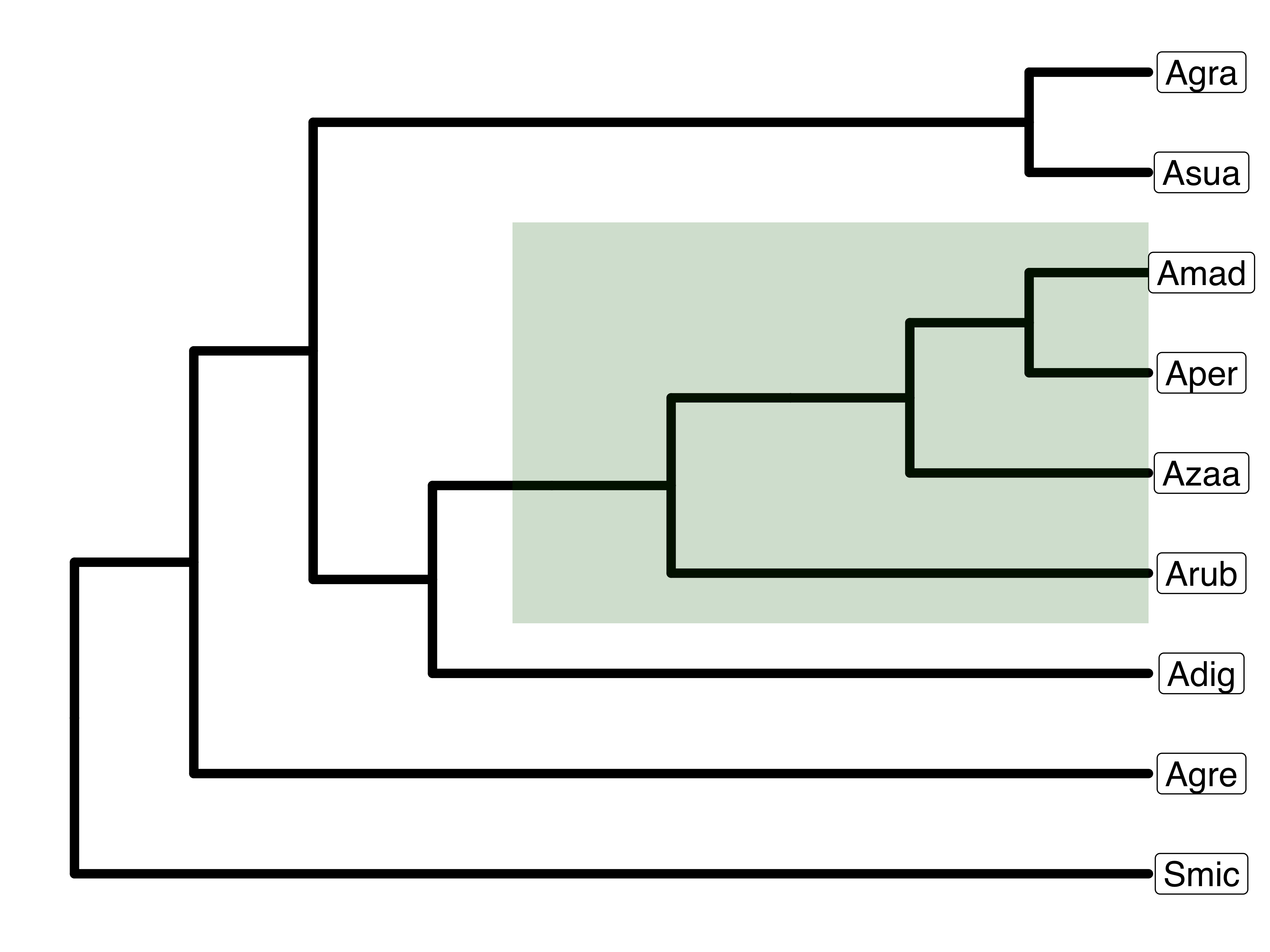}
}
\caption{Clustering via split-weight embedding recovers the correct evolutionary relationship in baobabs. Estimated gene trees from the baobabs group cluster around two consensus trees (341/363 genes around the center tree and 22/363 genes around the right tree) that represent the two correct possibilities created by the gene flow event in the phylogenetic network for the group (left) as estimated by the original study \cite{baobabs}. The placement of the (\textit{A.mad, A.per, A.zaa}) clade changes from being sister to clade (\textit{A.gra, A.sua}) in the center tree (clade highlighted in blue) to be sister of taxon \textit{A.rub} in the right tree (clade highlighted in green).}
\label{fig:baobabs}
\end{center}
\vskip -0.2in
\end{figure*}

After clustering of the sample of gene trees, the two resulting clusters are not balanced: 341 trees in cluster 1 and 22 trees in cluster 2 (Figure \ref{fig:pca} in the Appendix). This is expected given the evolutionary history of the baobabs group (Figure \ref{fig:baobabs} left). The original publication \cite{baobabs} identified one reticulation event (blue arrow) representing $11.8$\% gene flow. This means that we expect $11.8$\% of the genes to follow the blue arrow back to the root in their evolutionary history, and thus, most of the genes ($88.2$\%) will have a tree that puts clade ((\textit{A.mad}, \textit{A.ped}), \textit{A.zaa}) sister to (\textit{A.gra}, \textit{A.sua}) (Figure \ref{fig:baobabs} center; clade highlighted in blue) whereas a few genes ($11.8$\%) will place ((\textit{A.mad}, \textit{A.ped}), \textit{A.zaa}) sister to \textit{A.rub} (Figure \ref{fig:baobabs} right; clade highlighted in green). 

The accurate identification of the consensus trees for each cluster, in spite of estimation error and other biological processes in addition to reticulation, is an important result for phylogenetic inference. Estimating a phylogenetic network with 9 taxa (tree plus gene flow event in Figure \ref{fig:baobabs} left) could take up to two days of compute time depending on the method used \cite{Solis-Lemus2016-ux}. Clustering the input trees, building the consensus trees and reconstructing the network from the consensus trees cannot take more than a few minutes. Our work shows the promise of machine learning (unsupervised learning specifically) to aid in the estimation of phylogenetic trees and networks in a scalable manner.

\section{Discussion}
Here, we apply the first (to our knowledge) implementation of unsupervised learning in the space of phylogenetic trees
via the novel and powerful split-weight embedding.
The split-weight embedding is related to BHV space embedding \cite{billera2001geometry} in that it uses a vector of edge weights to represent a tree. The main difference is that the split-weight embedding is more sparse by the addition of zeroes for splits not present in the tree. This sparsity allows embedding vectors to be in the same Euclidean space (which is not true for BHV space), but it will impose computational challenges for large trees (more below).
Tropical geometric space \cite{monod2018tropical, lin2022tropical} or the hyperbolic embedding \cite{matsumoto2021novel, jiang2022learning} embed nodes in the trees as vectors that preserve phylogenetic similarity (such as pairwise distances among the leaves). While these spaces can enhance phylogenetic hierarchical structure better than Euclidean spaces, there are no simple implementations of embedding and distance algorithms for use among the evolutionary biology community, and we show here that the much simpler split-weight embedding preserves enough phylogenetic signal to provide biologically meaningful clustering.

Via extensive simulations, we show that the split-weight embedding is able to capture meaningful evolutionary relationships while keeping the simplicity of a standard Euclidean space.
Our implementation is able to cluster trees with different topologies, and even trees with the same topology, but different edge weights. As usual in machine learning applications, the larger the sample size (number of trees), the more accurately the different clusters were recovered.
On average, K-means was the desired choice of algorithm as it showed robust performance across sample size and number of taxa, yet for large trees (8 or 16 taxa), hierarchical clustering outperforms K-means in terms of running efficiency and accuracy. For the case of 4 taxa, GMM outperforms K-means when the sample size increased.

The bottleneck of our implementation is the curse of dimensionality. In its current version, we do not perform dimension reduction except for visualization purposes. The dimension of the split-weight embedded vector is given by the number of bipartitions which is $2^{n-1} - 1$ for $n$ taxa. The exponentially increasing nature of vectors may limit the application of split-weight embedding and incur significant computational cost. In addition, the embedded vector is highly sparse. For a tree of $n$ taxa, there are $2n-3$ edges, and thus, only $2n-3$ entries of the embedded vector will be different than zero. 
So, for example, for $n=8$ taxa, the embedded vector is 127-dimensional with 13 non-zero entries; for $n=16$ taxa, the embedded vector is 32767-dimensional with 29 non-zero entries. The field of phylogenetic trees deals with datasets of hundreds or thousands of taxa consistently.
Future work will involve the study of dimension reduction techniques, as well as compression, such as autoencoder models in order to improve the scalability and stability of our algorithms. 


Finally, here we utilize Densitree \cite{densitree} to obtain a representation of the consensus tree per cluster. We can, however, explore similar ideas to those in BHV space regarding the computation Fr\'echet sample means and Fr\'echet sample variances \cite{brown2020mean} which could set the foundation of classical statistical theory on the split-weight embedding space, for example, selective inference for the estimated clusters \cite{Chen2023}.


\section{Data and code availability}
 The baobabs dataset was made publicly available by the original publication \cite{baobabs} and can be accessed through the Dryad Digital Repository \url{http://doi.org/10.5061/dryad.mf1pp3r}. The split-weight embedding and unsupervised learning algorithms are implemented in the open source publicly available Julia package 
 available in the GitHub repository \url{https://github.com/solislemuslab/PhyloClustering.jl}.
 All the simulation and real data scripts for our work are available in the public GitHub repository \url{https://github.com/YiboK/PhyloClustering-scripts}.



\section{Acknowledgements}
This work was supported by the National Science Foundation [DEB-2144367 to CSL]. The authors thank Marianne Bj\"{o}rner and Reed Nelson for help with setting up tests, and Zhaoxing Wu for providing testing data.

\printbibliography

\newpage
\appendix

\section{Supplementary Figures}

\FloatBarrier
\subsection{Clustering of trees with different topologies}
\FloatBarrier

\begin{figure*}[!ht]
    \centering
    \includegraphics[scale=0.125]{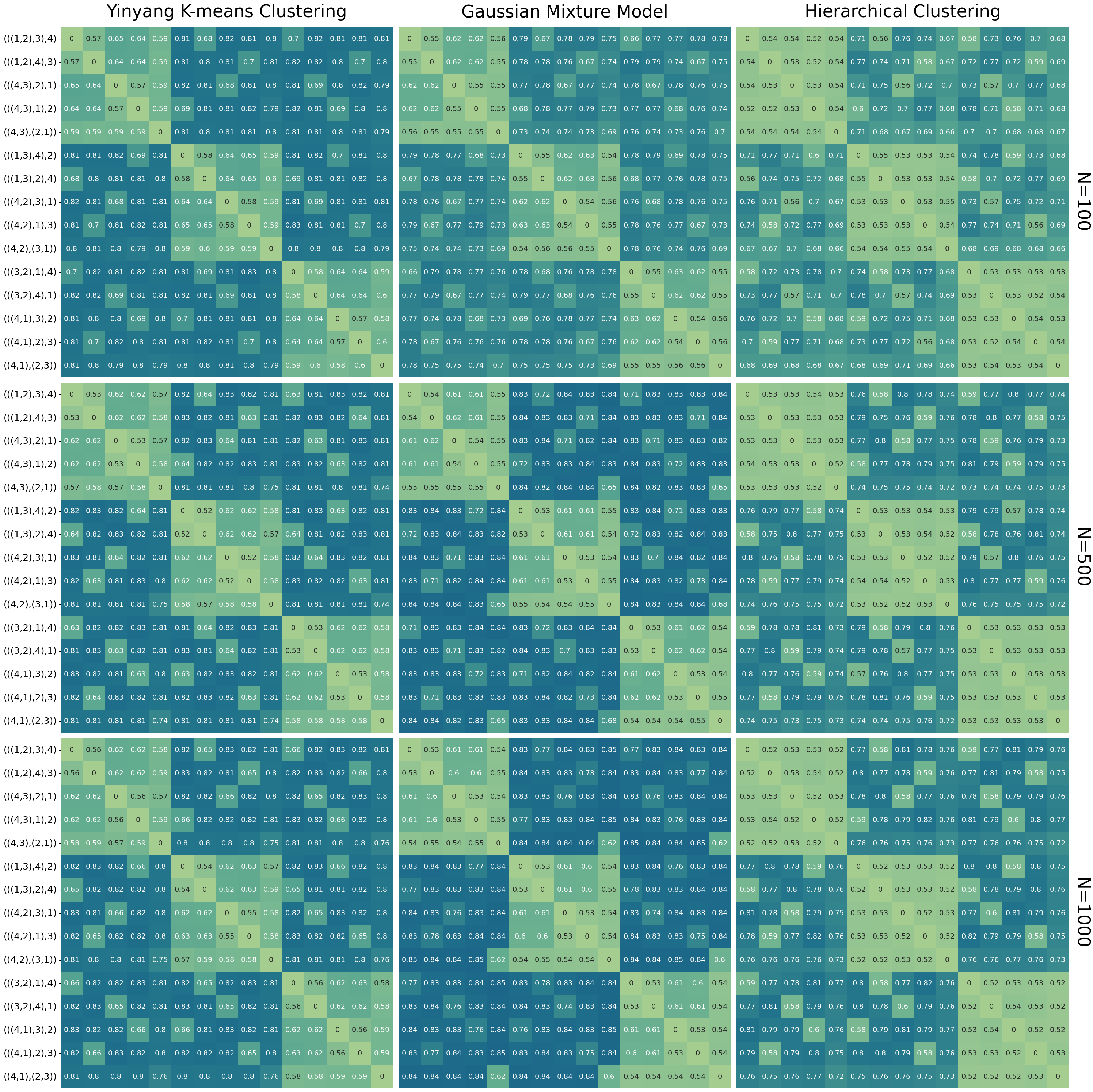}
    \includegraphics[scale=0.125]{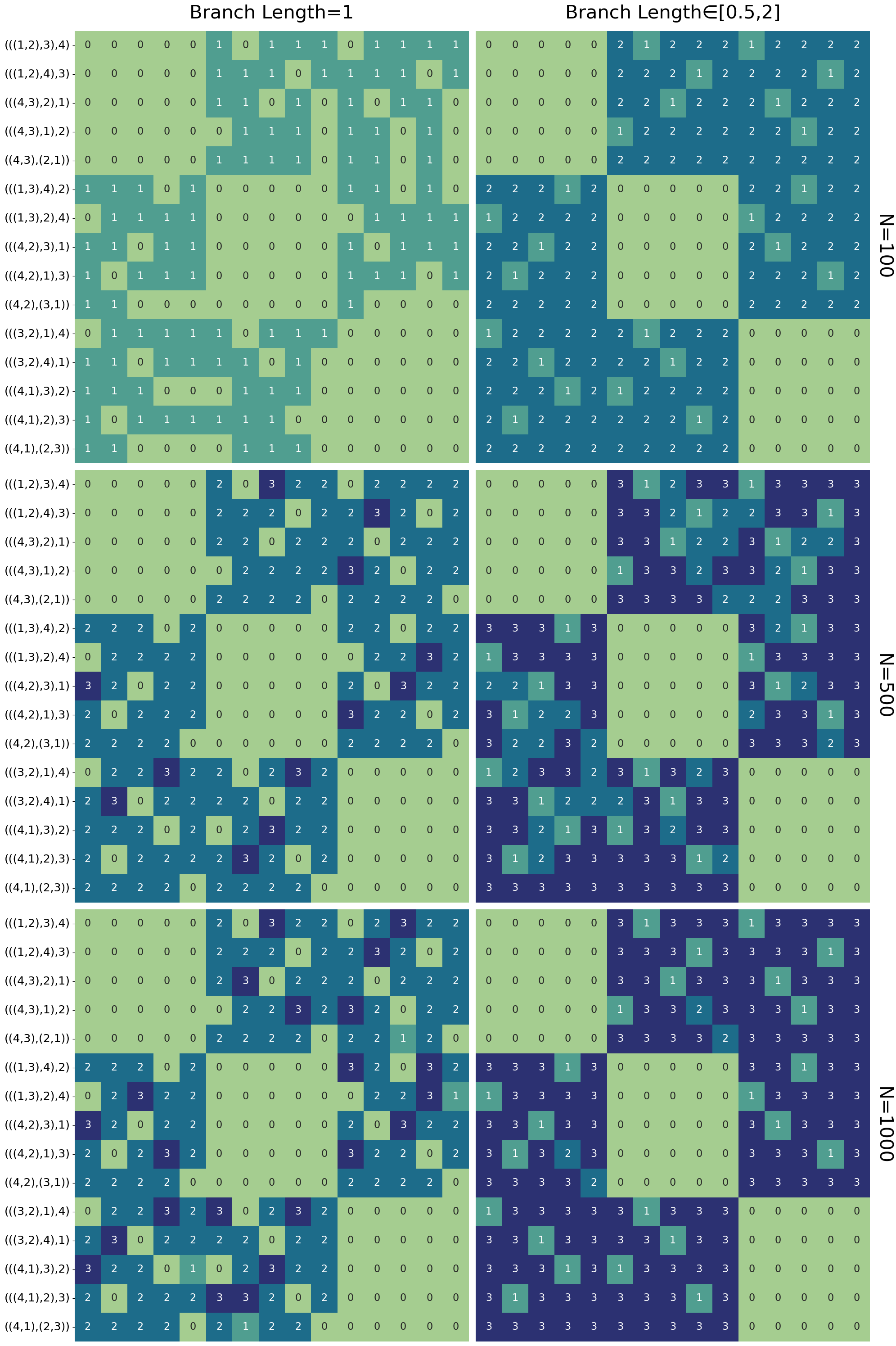}
    \caption{Left: Heatmaps of classificacion accuracy on the 15 trees with 4 taxa. Each panel corresponds to the classification accuracy of pairwise comparisons for two clusters originated on two trees (row and column) for one clustering method (columns: K-means, GMM and Hierarchical) and one sample size (rows: $N=100,500,1000$). Within each panel, trees are sorted based on their unrooted topology (first 5 rows correspond to split $12|34$; next 5 to the split $13|24$, and last 5 to the split $14|23$). Right: Heatmaps on the number of algorithms that reach $80$\% classification accuracy ($0,1,2,$ or $3$ algorithms) on edge weight strategy (columns: edge weights equal to $1$ or edge weights randomly chosen in $(0.5,2)$) and sample size (rows: $N=100,500,1000$).}
    \label{fig:4taxab}
\end{figure*}

\begin{figure*}[!ht]
    \centering
    \includegraphics[scale=0.13]{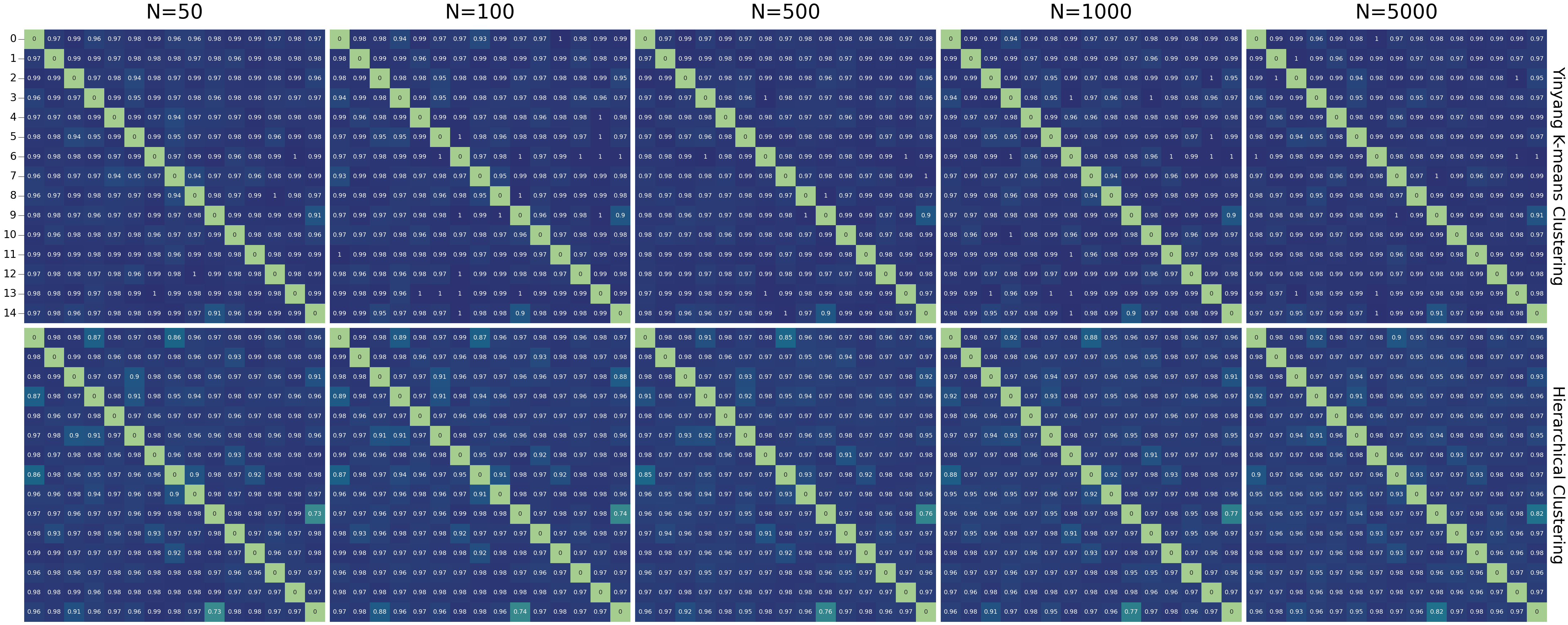}
    \caption{Heatmaps of classificacion accuracy on the 15 randomly generated trees with 8 taxa. Each panel corresponds to the classification accuracy of pairwise comparisons for two clusters originated on two trees (row and column) for one clustering method (rows: K-means and Hierarchical) and one sample size (columns: $N=50,100,500,1000,5000$).}
    \label{fig:8taxa}
\end{figure*}

\begin{figure*}[!ht]
    \centering
    \includegraphics[scale=0.13]{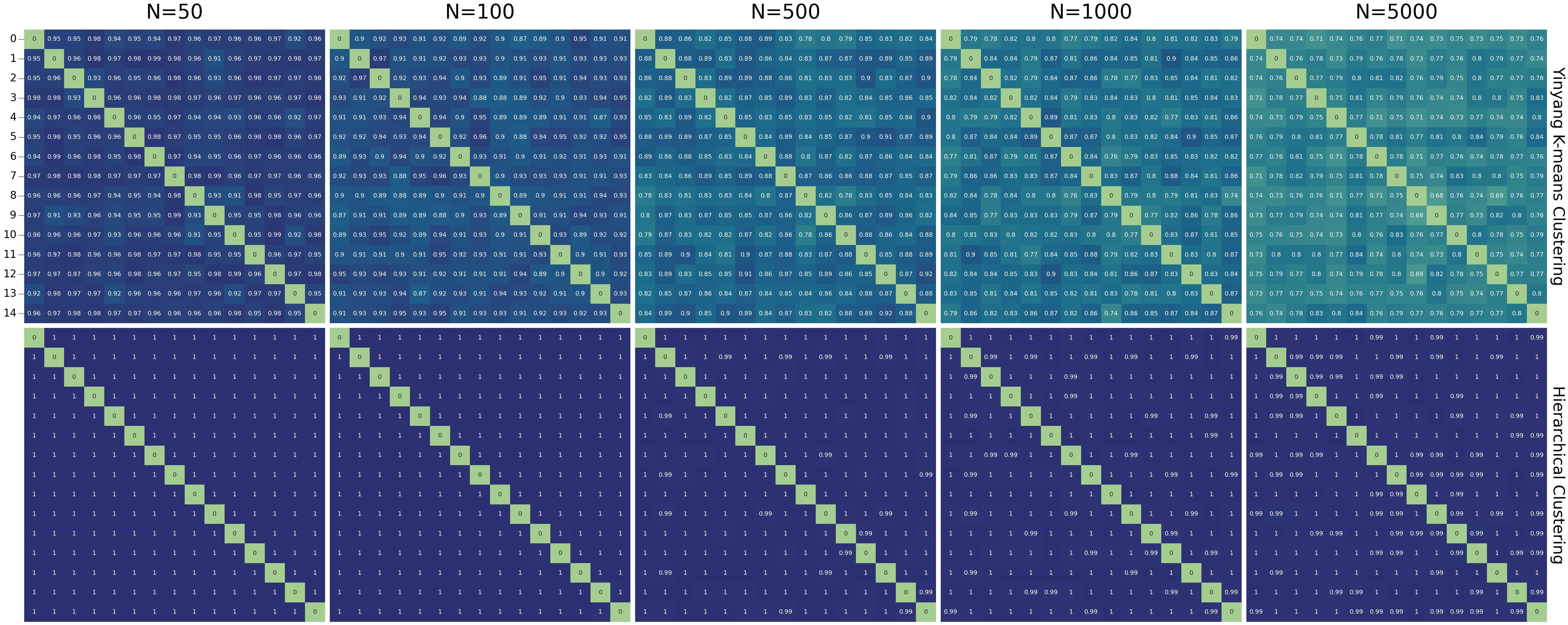}
    \caption{Heatmaps of classificacion accuracy on the 15 randomly generated trees with 16 taxa. Each panel corresponds to the classification accuracy of pairwise comparisons for two clusters originated on two trees (row and column) for one clustering method (rows: K-means and Hierarchical) and one sample size (columns: $N=50,100,500,1000,5000$).}
    \label{fig:16taxa}
\end{figure*}

\FloatBarrier
\subsection{Clustering of trees in NNI-neighborhoods}
\FloatBarrier

\begin{figure}[!ht]
    \centering
    \includegraphics[scale=0.15]{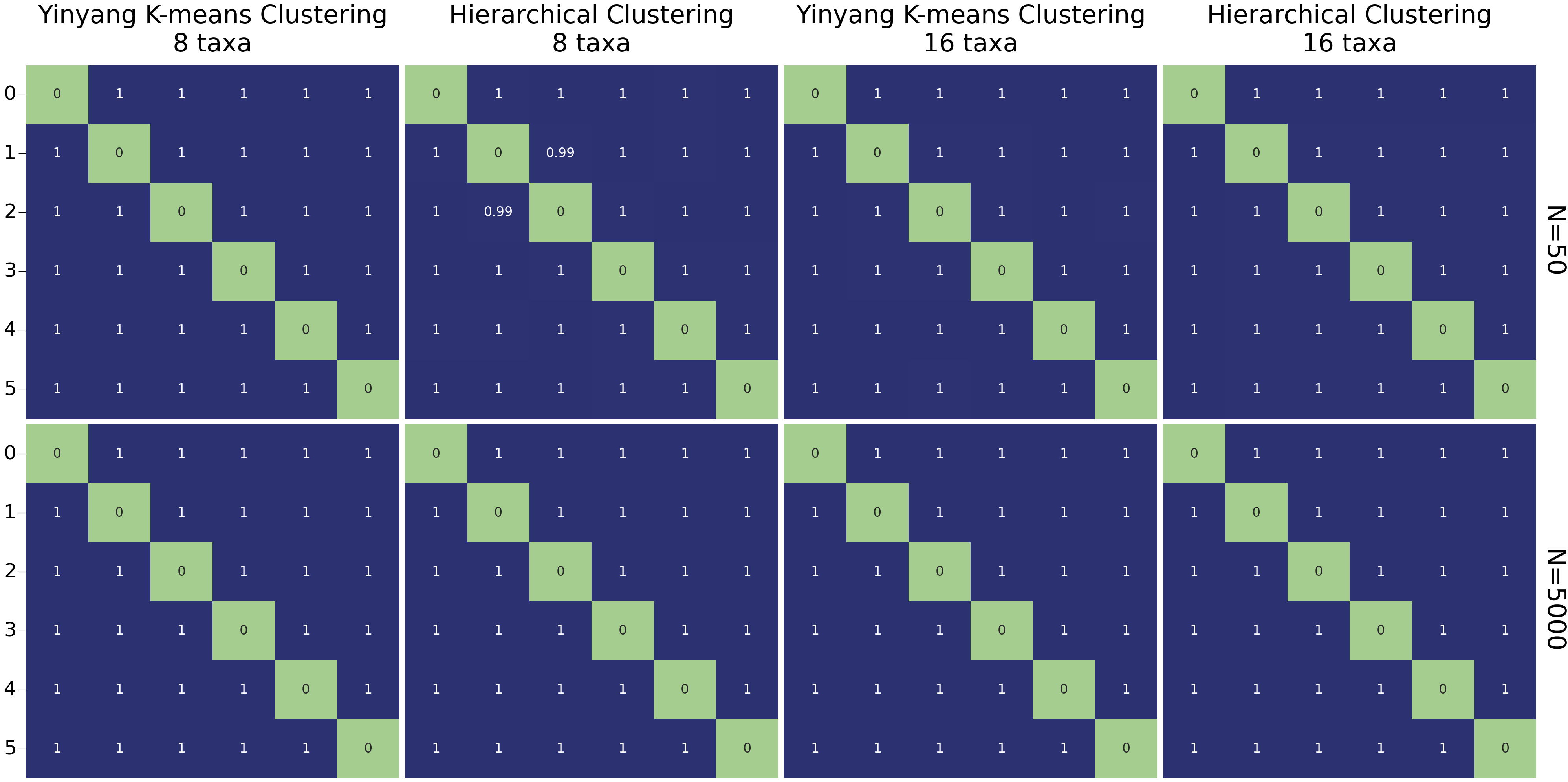}
    \caption{Heatmaps of classification accuracy on one randomly generated tree with 8 taxa (first two columns) or 16 taxa (second two columns) with 5 other trees obtained by $1,2,3,4$ or $5$ NNI moves on the original tree. Note "1" in the figure indicates 100\% accuracy. Rows correspond to sample size ($N=50$ trees on top, and $N=5000$ trees on bottom). Classificacion accuracy is high for both methods even for very similar trees (1 NNI move away, for example) which illustrates the preservation of phylogenetic signal by the split-weight embedding.}
    \label{fig:nni}
\end{figure}

\begin{figure*}[!ht]
    \centering
    \includegraphics[scale=0.15]{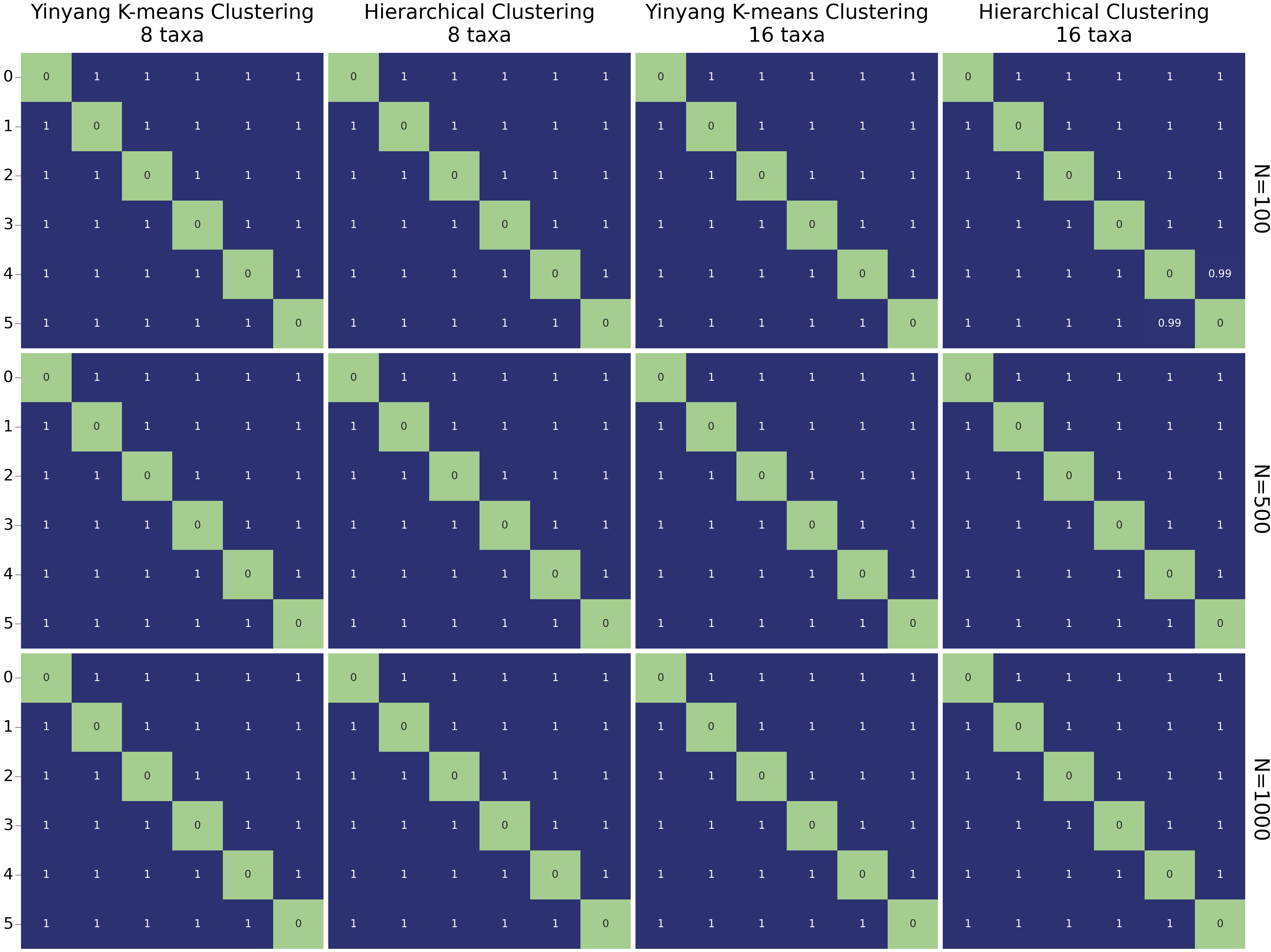}
    \caption{Heatmaps of classification accuracy on one randomly generated tree with 8 taxa (first two columns) or 16 taxa (second two columns) with 5 other trees obtained by $1,2,3,4$ or $5$ NNI moves on the original tree. Rows correspond to sample size ($N=100,500,1000$). Classificacion accuracy is high for both methods even for very similar trees (1 NNI move away, for example).}
    \label{fig:nnib}
\end{figure*}

Figure \ref{fig:nni} shows the classification accuracy of trees with 8 and 16 taxa, and their corresponding neighbor trees obtained by performing $1,2,3,4$ or $5$ NNI moves on the original tree. Despite the similarity of the trees under comparison, the methods are able to classify quite accurately clusters of trees originated from two similar trees. This implies that the split-weight embedding is able to preserve the necessary signal to classify phylogenetic trees even for closely related clusters. Figure \ref{fig:nnib} shows the same figure for $N=100,500,1000$ trees.

\FloatBarrier
\subsection{Clustering of trees with the same topology, but different edge weights}
\FloatBarrier

\begin{figure}[!ht]
    \centering
    \includegraphics[scale=0.16]{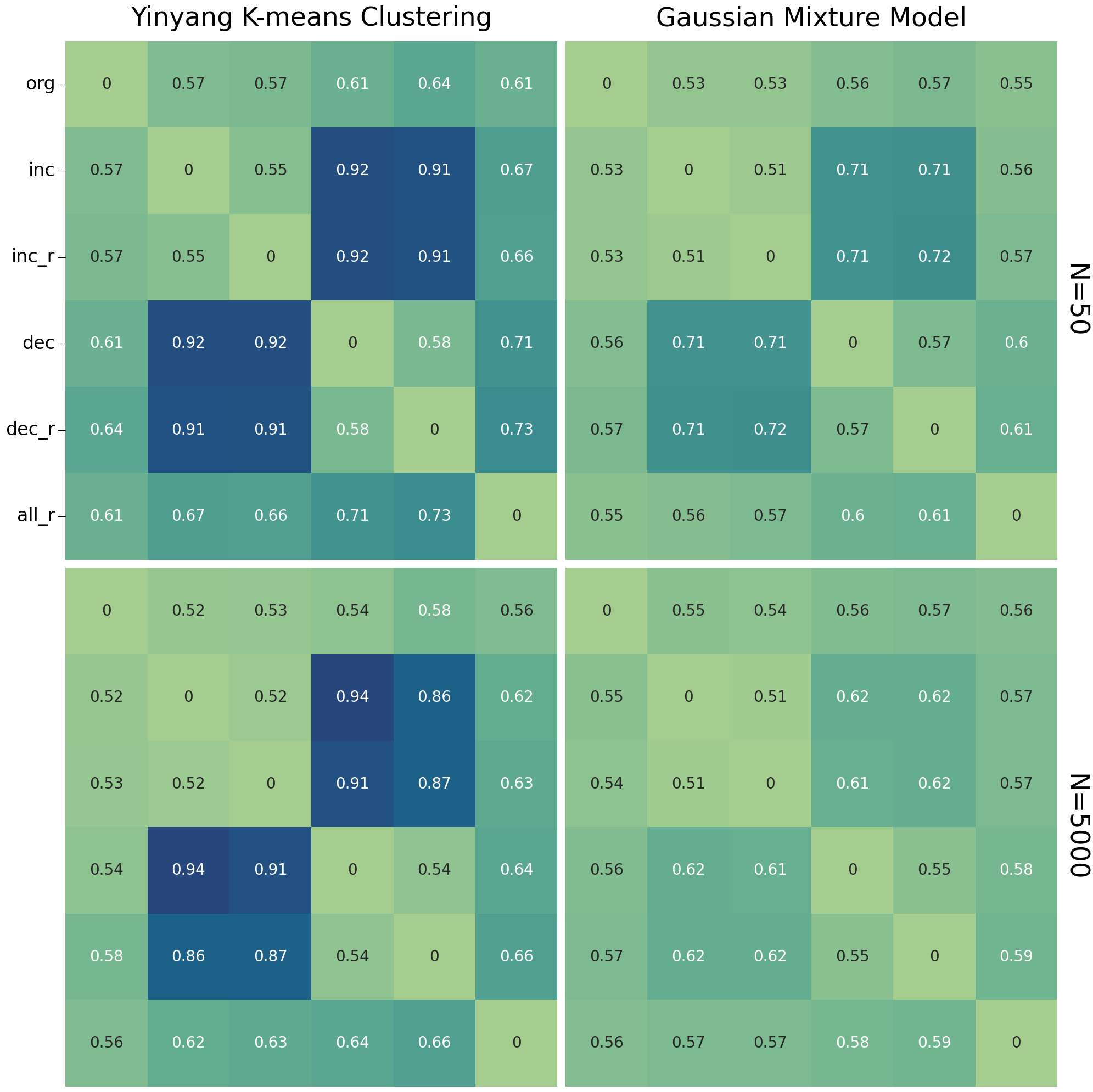}
    \caption{Heatmaps of classification accuracy on 4-taxon tree $(((4,3),2),1)$ by the two methods (columns: K-means and GMM) and sample size (rows: $N=50$, $N=5000$) by different set of edge weights. Clusters defined by the same tree topology can be accurately identified by K-means if edge weights are sufficiently different (\textit{inc} vs \textit{dec}, for example), but not by GMM.}
    \label{fig:bl1}
\end{figure}

\begin{figure}[!ht]
    \centering
    \includegraphics[scale=0.13]{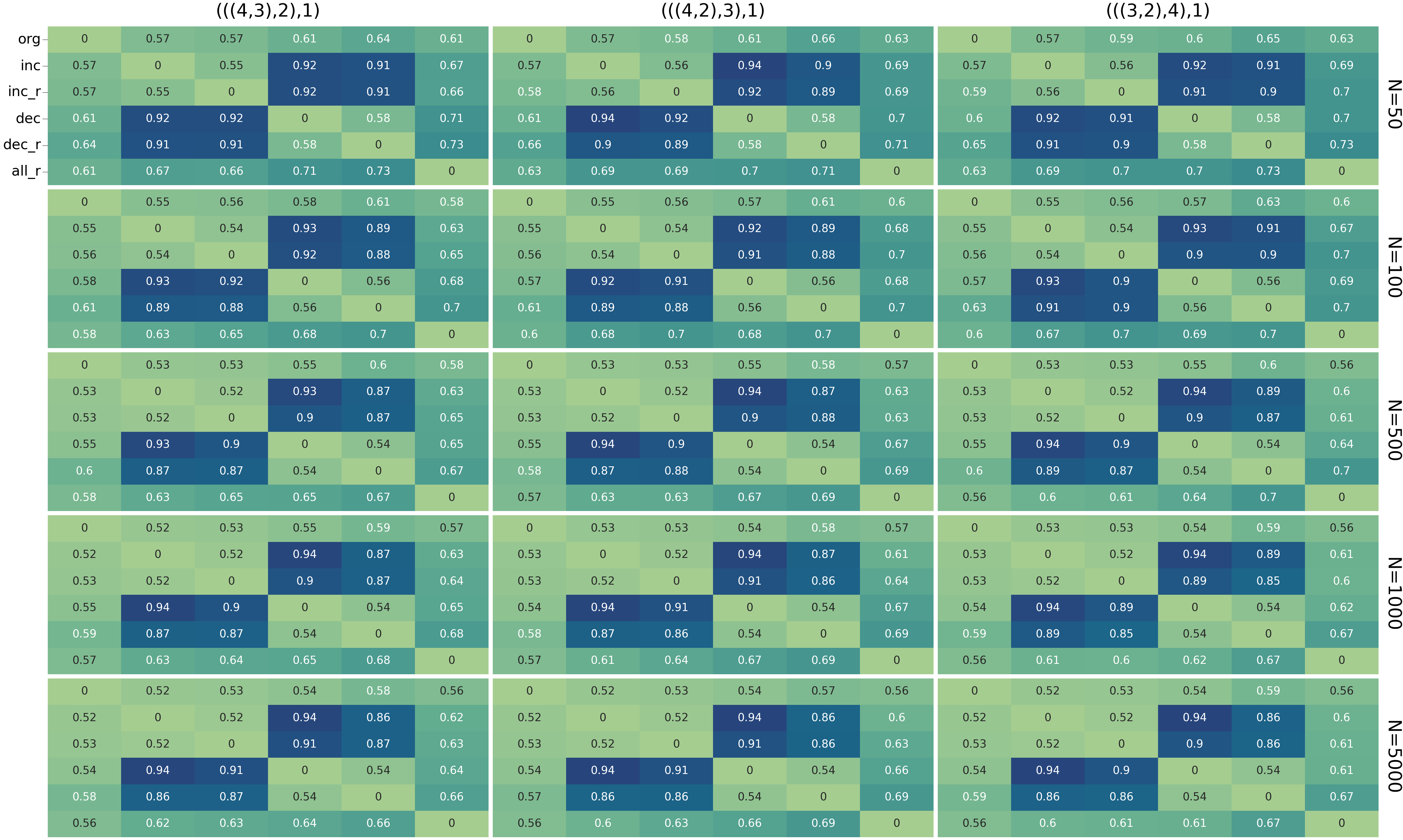}
    \caption{Heatmaps of K-means classification accuracy on three 4-taxon trees (columns) by sample size (rows: $N=50, 100, 500, 1000, 5000$) by different set of edge weights. Clusters defined by the same tree topology can be accurately identified if edge weights are sufficiently different (\textit{inc} vs \textit{dec}, for example).}
    \label{fig:bl2}
\end{figure}

\begin{figure}[!ht]
    \centering
    \includegraphics[scale=0.13]{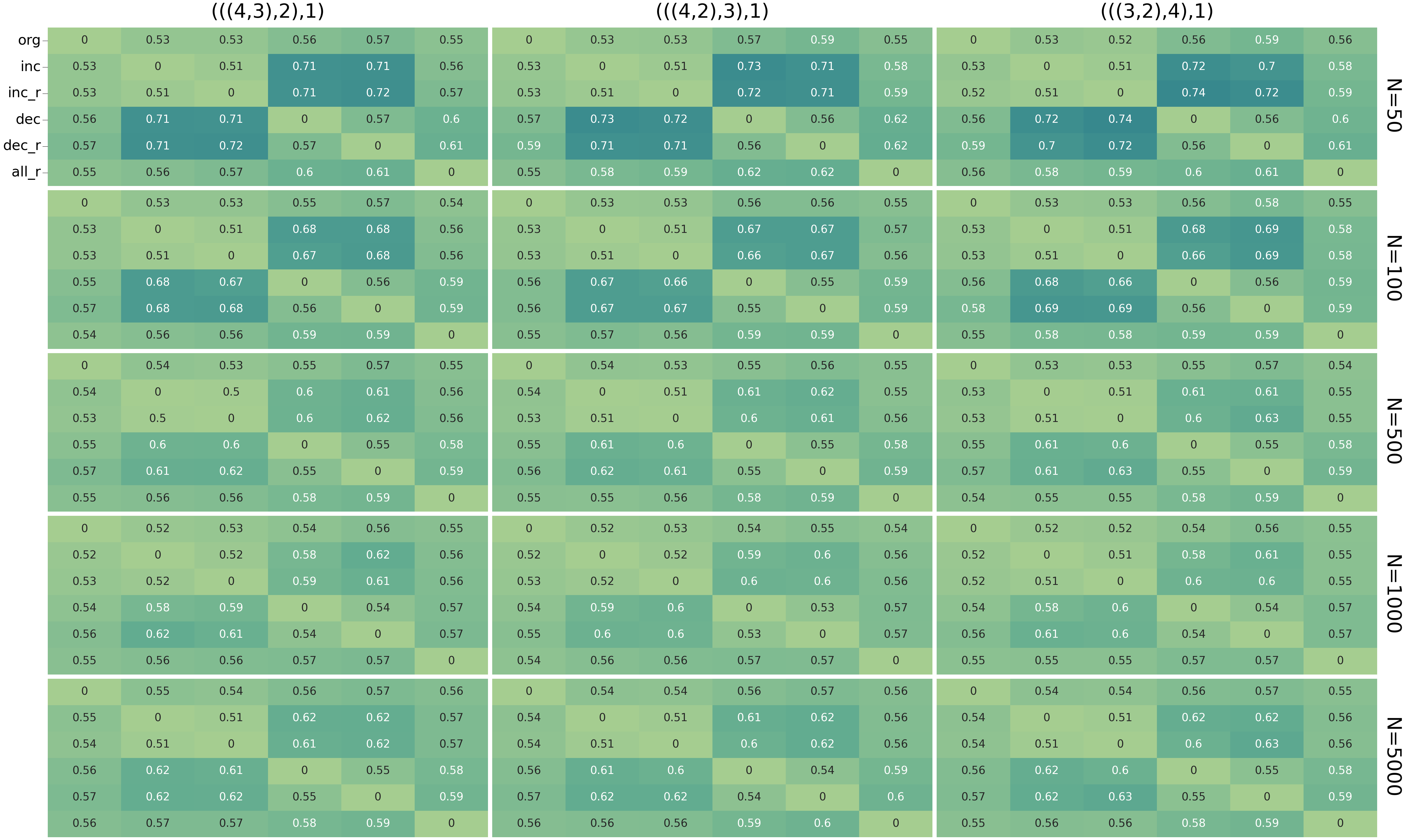}
    \caption{Heatmaps of GMM classification accuracy on three 4-taxon trees (columns) by sample size (rows: $N=50, 100, 500, 1000, 5000$) by different set of edge weights. Clusters defined by the same tree topology cannot be accurately identified compared to using K-means (Figure \ref{fig:bl2}).}
    \label{fig:bl3}
\end{figure}

While we showed that GMM can accurately identify tree clusters defined by different topologies (Figure \ref{fig:4taxa}), it appears that this algorithm does not have enough sensitivity to identify clusters originated from the same tree topology (Figure \ref{fig:bl1}). K-means, on the other hand, is able to identify such clusters as long as the edge weights are sufficiently different (\textit{dec} vs \textit{inc}, for example). Figures \ref{fig:bl2} and \ref{fig:bl3} for other tree topologies with similar conclusion.

\FloatBarrier
\subsection{Clustering of trees with imbalanced number of member}
\FloatBarrier

Table \ref{tab:imbalanced} shows the results.
We can observe that when $n=4$, there algorithms have lower ARI since the two groups in test are closter with each other so they are harder for the algorithms to distinguish. On the other hand, when $n=8$, the two groups are more seperated so they are easier to cluster.

\begin{table}[!h]
\centering
    \begin{subtable}{.5\linewidth}
    \caption*{ARI for $n = 4$}
    \centering
        \begin{tabular}{ |c|c|c|c| } \hline
         $2N$ & K-means & GMM & hierarchical \\
        \hline
        1100 & 0.326 & 0.570 & 0.376 \\ \hline
        6000 & 0.406 & 0.672 & 0.399 \\
        \hline
        \end{tabular} 
    \end{subtable}%
    \begin{subtable}{.5\linewidth}
    \caption*{ARI for $n = 8$}
    \centering
        \begin{tabular}{ |c|c|c|c| } \hline
         $2N$ & K-means & GMM & hierarchical \\
        \hline
        1100 & 0.967 & 0 & 0.908 \\ \hline
        6000 & 0.952 & 0 & 0.921 \\
        \hline
        \end{tabular} 
    \end{subtable}%
\caption{Left: The ARIs of clustering algorithms with $n=4$. Right: The ARIs of clustering algorithms with $n=8$. GMM does not work when $n \geq 8$, so it has ARI = 0 in both cases.}
\label{tab:imbalanced}
\end{table}

\FloatBarrier
\subsection{Baobabs data}
\FloatBarrier

\begin{figure}[!ht]
\vskip 0.2in
\begin{center}
\centerline{
\includegraphics[scale=0.03]{baobabs-c2.png}
\includegraphics[scale=0.03]{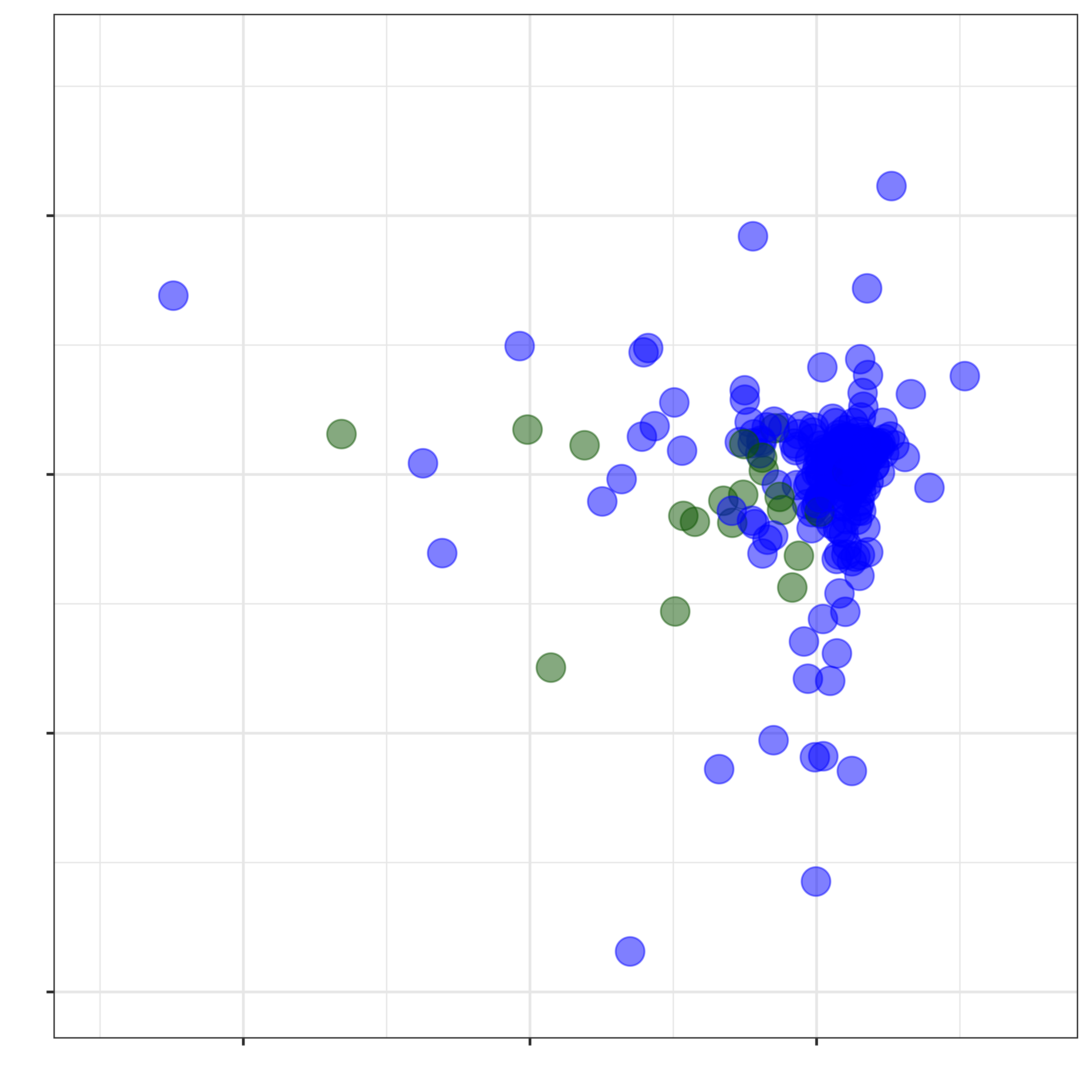}
\includegraphics[scale=0.03]{baobabs-c1.png}
}
\caption{PCA plot for the clustering of baobabs phylogenetic trees (center). The two clusters (blue and green) are highly imbalanced (341/363 vs 22/363), but these partition agrees with the evolutionary history of the group (Figure \ref{fig:baobabs} in the main text) which involves one gene flow event for $11.2$\% of the genes \cite{baobabs}. Left and right trees represent the consensus tree created by Densitree \cite{densitree} for each cluster.}
\label{fig:pca}
\end{center}
\vskip -0.2in
\end{figure}

\end{document}